\documentclass{perso}
\usepackage{cite}
\usepackage{graphicx}
\usepackage{amssymb}
\usepackage{amsmath}
\usepackage{multicol}

\shorttitle{How to guess the inter magnetic bubble  ETC.}
\title{How to guess the inter magnetic bubble potential by using a simple perceptron ?}
\author{Stephane Padovani} 
\institute{stephane.padovani@wanadoo.fr}


\begin{document}
\maketitle
\begin{abstract}
It is shown that magnetic bubble films behaviour can be described by using a 2~D super-Ising hamiltonian. Calculated hysteresis curves and magnetic domain patterns are successfully compared with experimental results taken in literature. The reciprocal problem of finding paramaters of the super-Ising model to reproduce computed or experimental magnetic domain pictures is solved by using a perceptron neural network.
\end{abstract}
Thin films with perpendicular magnetization show widespread technological applications,
particularly as high density storage devices. Thus, our understanding, at a fundamental level, of the way the magnetic domains are organized in such films is extremely important. This letter is focussed on films having cylindrical shape bubble and labyrinthine magnetic domains. These films have been extensively studied in the seventies as they were considered as a possible support
for high density magnetic recording, essentially through the possibility of manipulating
bubbles \cite{Eschenfelder}. There is nowadays a renewed interest for bubble and
stripe domains \cite{Hehn,Masson}, because of the appearance of new techniques,
like Magnetic Force Microscopy (MFM) or Secondary Electron Microscopy with Polarization Analysis (SEMPA), showing these domain geometries at the nanometer scale. The theoretical
models used up to now to describe these domains were developped 30
years ago \cite{Kooy,Cape,Thiele}, and consist of a comparison of the energies of domains having
different ideal geometries. Such an approach is the simplest one can imagine but provides a limited understanding of the mechanism of domain formation. More elaborated
descriptions based on linearized micromagnetic equations \cite{Muller} or Ginzburg-Landau
formalism \cite{Garel} have been proposed. These approaches describe qualitatively
the transition from bubble to stripe structure, but they do not allow quantitative
comparison with experimental results.\\
\indent
It will be explained in a first part that bubble films behave as
an Ising system with
long-range interaction (super-Ising) between bubbles considered as giant spins. The influence
of the interaction potential parameters (range and shape) on the domain geometry
and on the magnetization curve will be discussed. The comparison between
simulations and experimental results will help to determine the relevant parameters. In a second part, we solve the inverse problem i.e the determination of the interaction potential from a simulated or experimental magnetic map. The treatment proposed is a perceptron type neural network (NNT) searching for correct values of the potential using an error-correction learning rule \cite{Rosenblatt}. To our knowledge, it is the first time that neural network techniques are proposed to analyse magnetic domain pattern geometries.\\
\indent
The magnetic behaviour of bubble films can be understood starting from a magnetic monodomain situation. This state becomes unstable if an initially large perpendicular magnetic field is decreased to reach a critical value \( B_{N} \) (nucleation field).
In fact, for such a field, a situation with lower energy is obtained
if a bubble with a diameter $d_0$ is formed (critical bubble). The competition between the wall and magnetostatic energies determines $d_{0}$. The magnetostatic energy favours the formation of domains and varies, in first approximation, with the volume ($V_{0} \propto d_{0}^{2}$) of the bubble. The wall energy, that does not favor
the presence of domains, is proportional to the bubble's surface (\(S_{0}\propto d_{0}\)). The wall thickness is assumed to be negligible \cite{Eschenfelder,Hehn,Masson,Kooy,Cape,Thiele}. The characteristic length  \( l=\sigma _{w}/4\pi M^{2}_{s} \) reflects the balance between the wall energy
and the magnetostatic energy,  \( \sigma _{w} \)
is the wall energy density and \( M_{s} \) is the saturation magnetization
of the material. Typically, for cobalt or iron-palladium alloy, \( l \) is
around 10 nm. This length allows us to treat a non-dimensionnal problem by choosing $l$, $4\pi M_{S}$ and $4\pi M_{S}^{2}hd_{0}^{2}$ as respectively length, field and energy units.
 The function \( d_{0}\left( h\right)  \) has
been tabulated \cite{Eschenfelder,Cape} and can be approached by $\widetilde{d_{0}}=0.15\widetilde{h}+3.86$ for $\widetilde{h} \lesssim 50 $ ; the tilded quantities are non-dimensionnal variables.\\
\indent
We have considered up to now an isolated bubble.
The particular case of
two {\em down} magnetized critical bubbles in an {\em up} magnetized film (see fig.~1) is the following. Either the two bubbles are not in contact and they only interact via the dipolar field, the
wall energy is then \( 2\times \sigma _{W}S_{0} \), or the bubbles are
in contact and the wall energy becomes \( \sigma _{W}(2\times S_{0}-S_{1}) \), \( S_{1} \) being the contact surface between the two bubbles. These two cases
can be summarized in a single model on a square lattice in which each cell
i, of size \( d_{0}\times d_{0} \) is given the value of \( \eta _{i}=\pm 1 \), corresponding to {\em up} and {\em down} magnetization.
Then the interaction energy between two cells reads:
\begin{equation}
\label{2}
\widetilde{E}^{int}_{i,j}=\frac{1-\eta _{i}\eta _{j}}{2}\frac{\delta _{j,i\pm 1}}{\widetilde{d}_{0}}+\left| \widetilde{E}^{dip}(\left| i-j\right| )\right| \eta _{i}\eta _{j}
\end{equation}
 where the first term is the wall energy and the second term the dipole energy.
The hamiltonian fully describing the interaction between cells is given by:
\begin{equation}
\label{3}
\widetilde{\begin{mathcal}H\end{mathcal}}=-\frac{1}{2}\sum _{i,j}\widetilde{V}_{i,j}\eta _{i}\eta _{j}-\sum _{i}\eta _{i}\widetilde{B}
\end{equation}
with $\widetilde{V}_{i,j}=\delta _{j,i \pm 1}/2 \widetilde{d}_{0}-\left| \widetilde{E}^{dip} \left( \left|i-j \right| \right) \right|$ if $i\neq j$ and $\widetilde{V}_{i,i}=0$.
The second term accounts for the Zeeman coupling between each bubble's momentum and
the applied magnetic field \( B \).
The hamiltonian (\ref{3}) is formally identical to a super-Ising one. Similar analysis, using the competition between wall and dipolar energies, were previously presented in \cite{Kirby,Lyb} but it is the first time that bubble magnetic films are shown as 2~D super-Ising systems. It can be noticed that a similar hamiltonian could be written to describe perpendicularly magnetized nanoscale dot arrays \cite{Aign}.
\begin{figure}[t]
{\par\centering \label{fig. 1}\resizebox*{0.6\columnwidth}{!}{\includegraphics{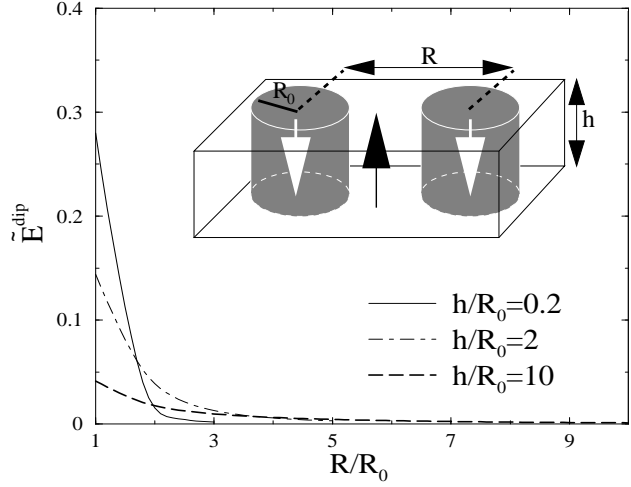}} \par}
\caption{Interaction dipolar energy between two bubbles of same sign as a function of their separation distance $R$ for different film thicknesses. The inset shows two {\em down} cylindar shape bubbles in an {\em up} magnetized film.
}
\end{figure}
\\
\indent If the dipolar field of a square  cell is approximated by the one of a cylindar of radius \( R_{0}=d_{0}/2 \), $\widetilde{E}^{dip}$ reads:
\begin{equation}
\label{5}
\widetilde{E}^{dip}\left( \left| i-j\right| \right) =\frac{\pi R_{0}}{2h}\int ^{+\infty }_{0}d\epsilon \left( \frac{J_{1}(\epsilon )}{\epsilon }\right) ^{2} J_{0}\left( \epsilon \left| i-j\right| \right) \left( 1-e^{-\epsilon h/R_{0}}\right)
\end{equation}
where $J_{0}$ and $J_{1}$ are Bessel functions of order respectively 0 and 1.
The dipole term is short ranged even
if its range increases with the film thickness (see fig.~1).
For simplification of the forthcoming discussion, the potential \( \widetilde{V} \) is approximated by:
\begin{equation}
\label{6}
\widetilde{V}(|i-j|)=\left\{ \begin{array}{ccc}
\widetilde{V}_{+} & \text{if} & 0<\left| i-j\right| \leq \widetilde{R}_{+}\\
\widetilde{V}_{-} & \text{if} & \widetilde{R}_{+}<\left| i-j\right| \leq \widetilde{R}_{-}\\
\end{array}\right.
\end{equation}
where \( V_{+} \) is the sum of the dipolar and the wall energy between
first neighbours, \( V_{-} \) is the dipolar energy, \( R_{+} \) the distance to the first neighbours and \( R_{-} \) is the range of
the potential.
For the cobalt films studied in \cite{Hehn}, with thickness \( \widetilde{h}=50 \) and \( \widetilde{d_{0}}\simeq 10 \), \( \widetilde{V}_{+} \) and \( \widetilde{V}_{_{-}} \) can be respectively
fixed to 0.02 and -0.02, \( \widetilde{R}_{-}=4 \) and \( \widetilde{R}_{+}=\sqrt{2} \).
This choice of $\widetilde{R}_{+}$ is equivalent to consider also the second neighbours
of a cell as nearest neighbours. This allows the model to be more isotropic and
attenuates the lattice effects.\\
\indent With $\widetilde{\Delta}_{i}=\sum _{k} \widetilde{V}\left( k\right) \eta _{i+k}+\widetilde{B}$, the average
momentum of a cell i, for a given environment $env_{i}$, is in the canonical ensemble:
\begin{equation}
\label{7}
\left\langle \eta_{i}\right\rangle_{env_{i}} =\tanh \left( \frac{4 \pi M_{s}^2 d_{0}^{2}h}{k_{B}T} \widetilde{\Delta} _{i}\right)
\end{equation}
Since the ratio \( V_{+}/k_{B}\approx 10^{5}K \), in a reasonnable
\( T \) range, one can let \( T \) go to \( 0^{+} \) in (\ref{7}) and let the lattice follow the rule:
\begin{equation}
\label{9}
\eta _{i}=\left\{ \begin{array}{ccc}
sign\left( \widetilde{\Delta}_{i}\right)  & \text{if} & \widetilde{\Delta}_{i}\neq 0\\
\text{random choice} & \text{if} & \widetilde{\Delta}_{i}=0
\end{array}\right.
\end{equation}
\begin{figure}[t]
\begin{multicols}{2}
\vspace*{-0.27cm}
{\par\resizebox*{\columnwidth}{!}{\includegraphics{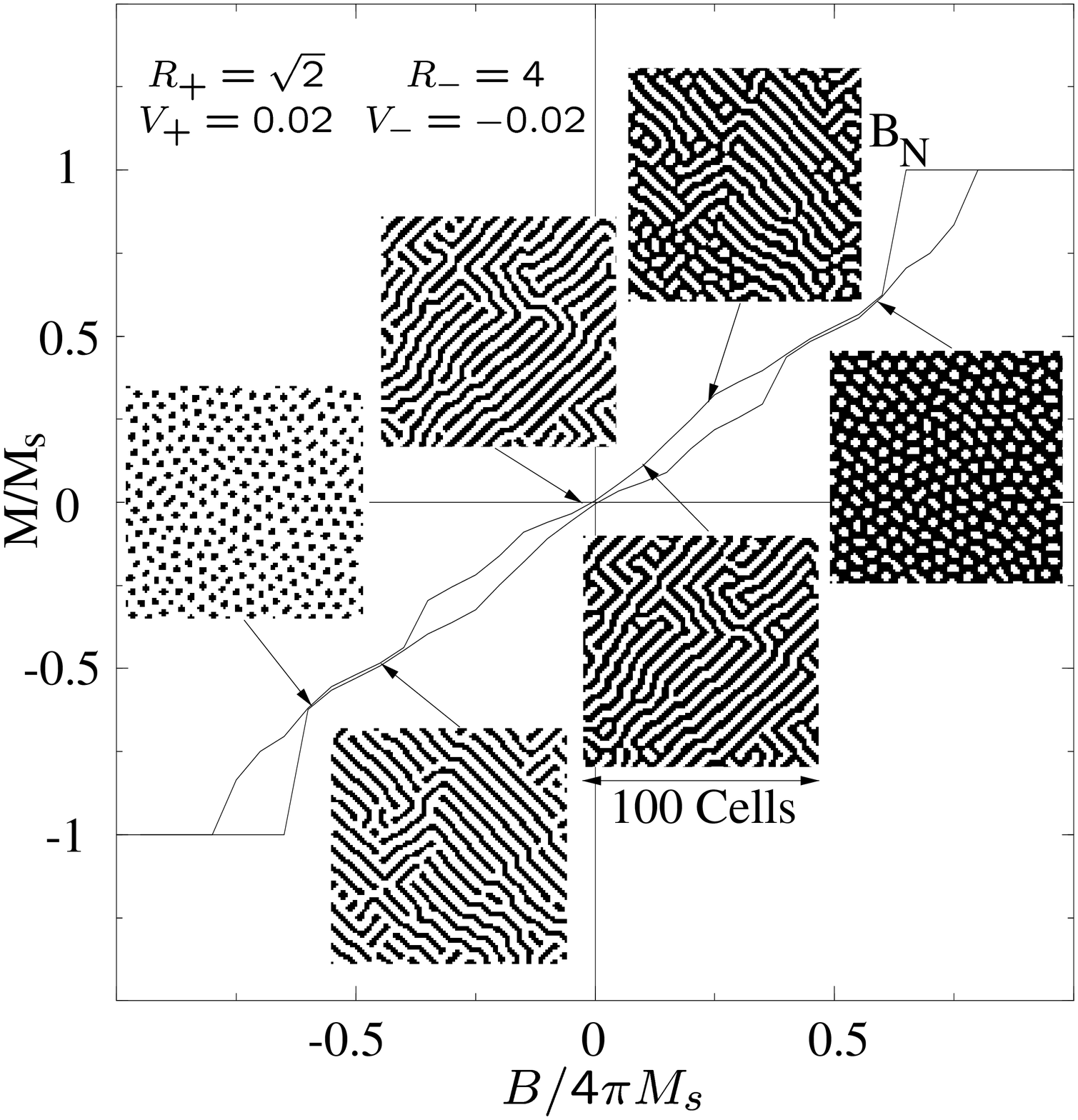}} \par}
\vspace*{-0.3cm}
{\par{\text{Fig. 2}} \par}
{\par\centering \resizebox*{1\columnwidth}{!}{\includegraphics{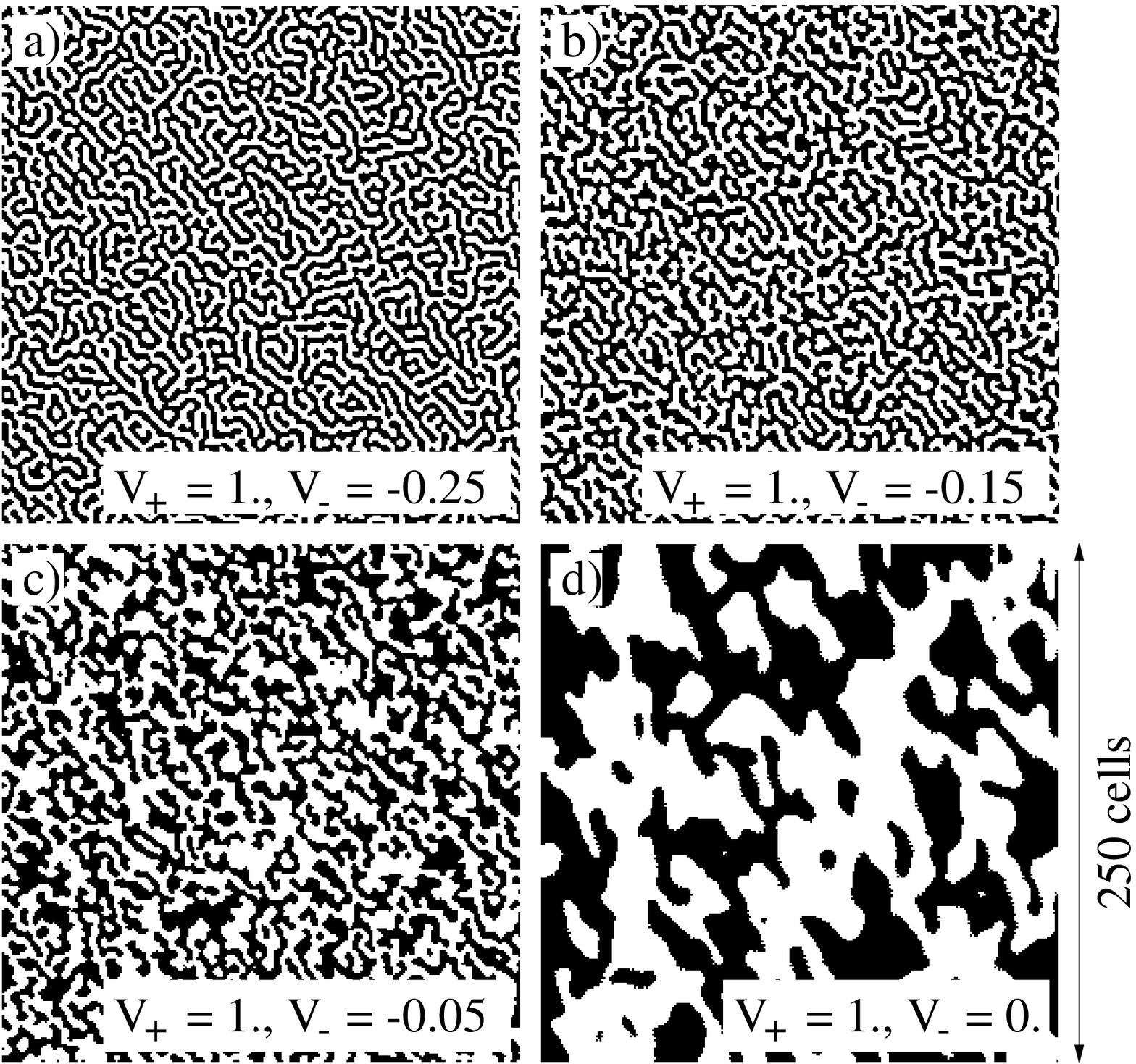}} \par}
\vspace*{0.25cm}
{\par{\text{Fig. 3}} \par}
\end{multicols}
\caption{Simulated magnetization curve. The computed patterns (100 $\times$ 100 cells) show the domain evolution at different points of the hysteresis curve when the field is decreased, starting from the saturated {\em up} state. Black and white parts correspond to {\em up} and {\em down} domains.}
\caption{Computed patterns in the demagnetized state for different potentiel values, $R_{+}=\sqrt{2}$ and $R_{-}=4$. Black and white parts correspond respectively to {\em up} and {\em down} domains.}
\end{figure}
The case $\widetilde{\Delta}_{i}=0$ corresponds to $\eta _{i}$ equals $\pm 1$ being equiprobable. This is crucial since decreasing \( \widetilde{B} \) from the monodomain state (all $\eta_i=+1$), the magnetization starts to flip for \( \widetilde{B}_{N}=-\sum _{k}\widetilde{V}(k) \), i.e. $\widetilde{\Delta}_{i}=0$.
Without dipolar field, the magnetization sharply flips for \( B_{N}<0 \)
with no domain nucleation (in this case \( B_{N} \) is similar to a coercitive
field). With dipolar field, \( B_{N} \) is possibly
positive and, in this case, the magnetization begins to decrease by creating domains: \( B_{N} \) is the nucleation field.\\
\indent
The hysteresis curve shown in fig.~2 is calculated for the parameters defined above. The similarity with the
experimental results from \cite{Hehn} is striking. Our computed value of \( B_{N}\simeq 10 \)
kOe (\( M_{s}=1480 \) \( erg.cm^{-3} \) for cobalt) is in good agreement with
regard to the experimental value ($\simeq$ 12 kOe).
The behaviour described by \cite{Kooy,Cape,Thiele} is found in our simulations.
When the field is decreased starting from the saturated {\em up} state (see fig.~2):  {\em down} bubbles
 first appear, decreasing the field further, bubbles are elliptically deformed (stripe out), then coalesce to form a maze structure, the domain walls move progressively making {\em up} domain size decreases, and finally {\em up} bubbles appear. We may emphasize that the whole evolution is completely described using only the hamiltonian (\ref{3}) with
no further hypothesis.\\
\indent
Domains in a demagnetized state for different potential values are shown in fig.~3. The $V_{+}$ part of the potential is arbitrarily fixed to unity since the calculation is performed under zero field. The demagnetized state is obtained starting from a full random initialization. This corresponds physically to demagnetize the film through annealing it at an infinite
temperature, and to cool it down abruptly. We have tested that it is not necessary
to use a complex procedure of simulated annealing to reach the ground state
of the system.
Our domain patterns in fig.~3 can be compared to the MFM images obtained by V. Gehanno {\it et al.} \cite{Masson}, for different film thicknesses. The domain geometry evolves from a maze type structure like our fig.~3a to a fig.~3c type domain
structure by decreasing the
film thickness from 50 nm to 5 nm, i.e by decreasing the \( V_{-} \) dipolar part of the potential. The pattern in fig.~3d fits well to SEMPA images obtained by R. Allenspach {\it et al.} \cite{Allenspach}
for ultra-thin cobalt films ($\sim$ 0.8 nm) on the (111) gold surface.
Since \( d_{0}\approx 50\: nm \), the image size in fig.~3a-d is around \( 10\: \mu m\: \times \: 10\: \mu m \), which gives a domain size: 2-5 $\mu$m for fig.~3d to 50-100 nm for fig.~3a-c, in good agreement with experimental data \cite{Masson,Allenspach}.
\begin{figure}[t]
\begin{multicols}{2}
{\par\resizebox*{1.\columnwidth}{!}{\includegraphics{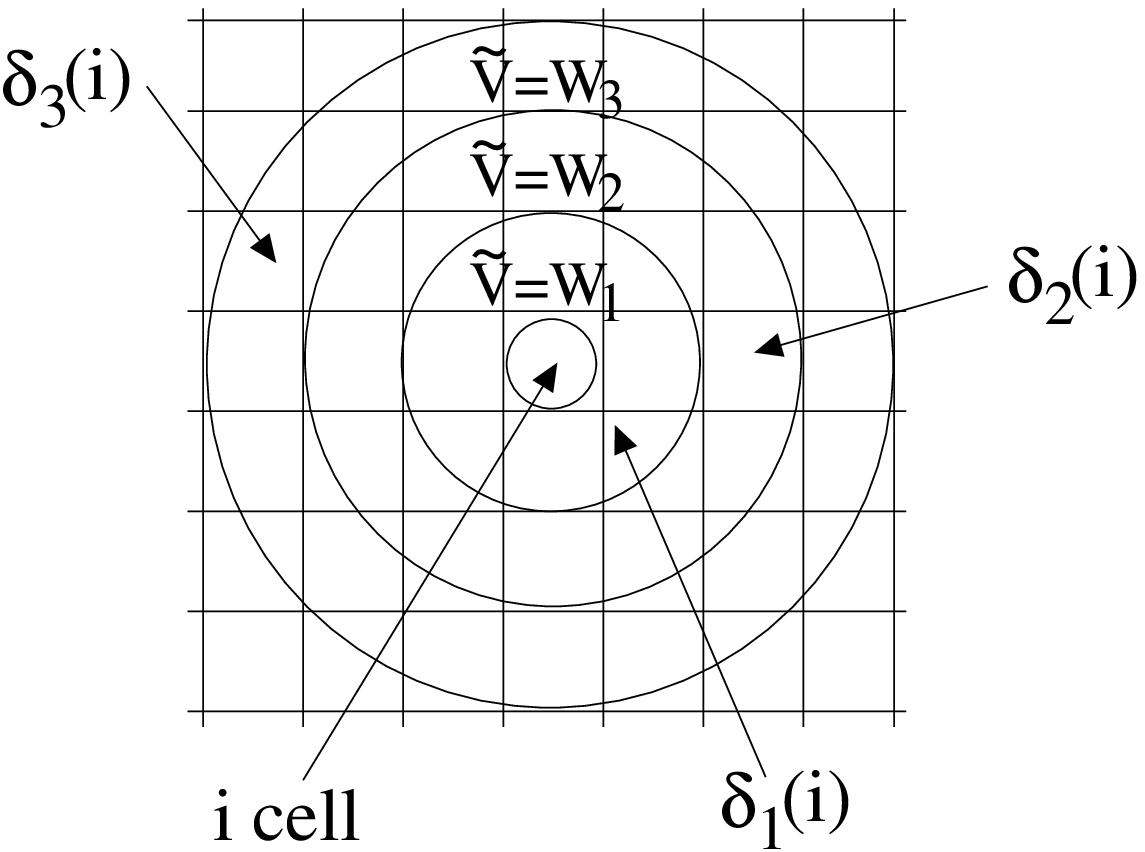}} \par} 
\vspace*{-0.1cm}
{\par\text{Fig. 4}\par}
{\par\resizebox*{1.\columnwidth}{!}{\includegraphics{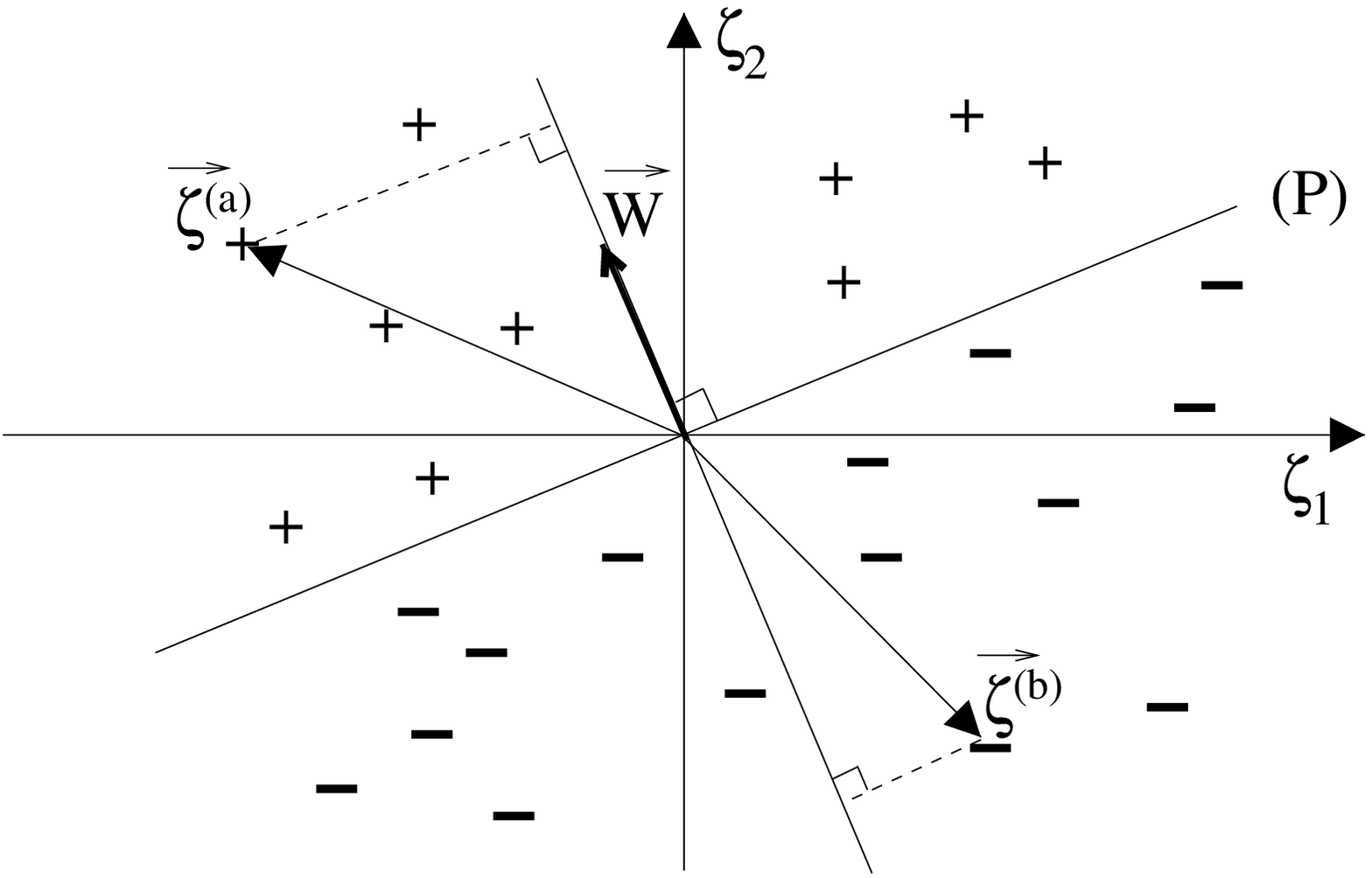}} \par}
\vspace*{0.55cm}
{\par\text{Fig. 5}\par}
\end{multicols}
\caption{Schematic representation of the first, second and third layers of the $i$ cell neighbours. The potential $\widetilde{V}$ is constant on each layer.}
\caption{$C_+$ and $C_-$ points are marked by respectively + and -. If a hyperplan $(P)$ separating $C_+$ from $C_-$ points is found, a vector $W$ perpendicular to $(P)$ verifies that the $\zeta ^{(i)}$ projection on $W$ is positive or negative depending $\zeta^{(i)}$ belongs respectively to $C_+$ or $C_-$ classes.}
\end{figure}

We now solve the inverse problem: can we from a given calculated or experimental domain pattern, infer the unknown potential $V$?
One needs to find the relation between the state of the cell \( i \)
and $env_{i}$. The \( i \) cell's environment can be described with the vector \( \overrightarrow{\zeta}^{(i)} \) whose the $q$ components are $\zeta _{q}^{(i)}=\sum _{p \in \delta_q(i)}\eta _{p}$, where the \( \delta_q(i) \) set states for the $i$ cell's q layer of neighbours (see fig.~4). We also define a vector $\overrightarrow{W}$ with components $W_q$ equal to $\widetilde{V}(|i-j|)$, the $j$ cell belonging to $\delta_q(i)$. We separate the \( \overrightarrow{\zeta}^{(i)} \) vectors into two classes $C_+$ and $C_-$ depending on \( \eta _{i} \) respectively
equals +1 or -1. If one excepts the marginal case where \( \widetilde{\Delta}_i=0 \)
(which is only true for \( B\neq B_{N} \)), the rule (\ref{9}) reads:
\begin{equation}
\label{11}
\left\{ \begin{array}{ccc}
\overrightarrow{\zeta }^{(i)}\in C_{+} & \iff & \left< \overrightarrow{\zeta}^{(i)},\overrightarrow{W}\right> + \widetilde{B} >0\\
\overrightarrow{\zeta }^{(i)}\in C_{-} & \iff & \left< \overrightarrow{\zeta} ^{(i)},\overrightarrow{W}\right> + \widetilde{B} <0
\end{array}\right.
\end{equation}
where $\left< \overrightarrow{\zeta }^{(i)}, \overrightarrow{W} \right> =\sum _q \zeta_q^{(i)} W_q=\sum_k \widetilde{V}_k\eta_{i+k}$ is the euclidian scalar product. In other words, one has to find a hyperplane $(P)$  separating \( C_{+} \) points from \( C_{-} \) points, thus a $\overrightarrow{W}$ vector perpendicular to $(P)$ will be solution of eq. \ref{11} (the case $\widetilde{B}=0$ is illustrated on fig.~5).\\
\indent
We start with a randomly chosen vector $\overrightarrow{W}^{(0)}$. For a given $i$ cell, the $\overrightarrow{\zeta}^{(i)}$ vector is calculated. If $ \left<\overrightarrow{W}^{(0)}, \overrightarrow{\zeta}^{(i)} \right> + \widetilde{B}$ and $\eta _i$ have the same sign, another cell is tested. In the other case, $\overrightarrow{W}^{(0)}$ is modified according to the rule
\begin{equation}
\label{12} 
\overrightarrow{W}^{(1)}=\overrightarrow{W}^{(0)}+\gamma \left[ \eta _i - sign \left( \left< \overrightarrow{W}^{(0)}, \overrightarrow{\zeta}^{(i)} \right> + \widetilde{B} \right)\right] \overrightarrow{\zeta }^{(i)}
\end{equation}
and so on until a vector $\overrightarrow{W}^{(n)}$ is found, verifying $\eta _i=sign\left( \left< \overrightarrow{W}^{(n)},\overrightarrow{\zeta}^{(i)}\right> + \widetilde{B} \right)$ for every $i$ cell. Finaly, the $\overrightarrow{W}^{(n)}$ components give the $\widetilde{V}$ potential we are looking for. 
\begin{figure}[t]
{\par\centering \label{fig. 4}\resizebox*{0.6\columnwidth}{!}{\includegraphics{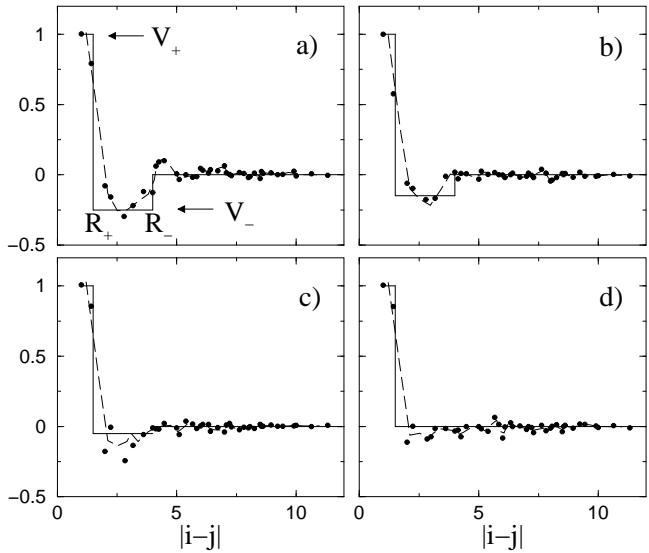}} \par}
\caption{Potentials corresponding respectively to a,b,c,d parts of fig.~3. The guessed potentials are represented by points, the ideal ones by solid lines and the dashed lines are the smoothed guessed potentials.}
\end{figure}
\\
\indent In fact, this algorithm follows closely the Rosenblatt's perceptron NNT \cite{Rosenblatt} in its simplest expression since only one neuron is needed. The $N$ components $\zeta^{(i)}_q$ are injected in the $N$ cells input layer of the perceptron. The neuron answer $\sigma$ is $sign\left( \left< \overrightarrow{W}^{(m)}, \overrightarrow{\zeta}^{(i)} \right> + \widetilde{B} \right)$, the $W^{(m)}_q$ being the synaptic coupling strengths between the $q$ cell of the input layer and the neuron. If the perceptron makes a mistake ($\eta_i \ne \sigma$), the synaptic couplings are corrected following eq. \ref{12}, this equation defining the error-correction learning rule of the perceptron. Papert and Minsky \cite{Papert} have shown that the perceptron NNT converges in a finite number of steps if $\gamma > 0$ and if the classification problem of the $\overrightarrow{\zeta}^{(i)}$ vectors between the two classes $C_+$ and $C_-$ is a linearly separable one\cite{rem}. In our case, the separability condition is verified by construction and for $\gamma=0.5$, a correct potential is found after only one scan of a 250 $\times$ 250 cells domain pattern.\\
\indent
The guessed potentials \( V/V(1) \) from demagnetized domain structures of fig.~3 are shown in fig.~6. The shape of the potential (positive between first neighbours and negative for other neighbours) as well as \( R_{-} \) and
\( R_{+} \) are estimated correctly. The determination of \(V_{-}/V_{+}\)
is less accurate because of statistical fluctuations. These latters have been reduced here
by taking the mean value of the curve between neighbour points (dashed line).
We have only used demagnetized states because it is often
the only information that can be experimentally obtained, but one can use domain patterns corresponding to different values of \( B \)
as an input and statistical
fluctuations will be strongly reduced. Nevertheless, the above results
show that even using a single image, and such a simple perceptron algorithm, it is possible to obtain
probing results.\\ 
\indent
In summary, magnetic bubble films can be described using a 2~D super-Ising hamiltonian. The potential allowing to fit an experimental or calculated domain pattern can be guessed by a perceptron neural network. The universality of the super-Ising model and of the bubble and maze domains \cite{Seul} suggests this approach could be exploited in numerous other problems as demixion processes \cite{demi}, Turing mechanisms \cite{tur} or surface gases \cite{Cu} to analyse experimental images.

\stars

I am very grateful to F. Bardou for fruitful discussions.


\begin{thebibliography}{99}
\bibitem{rem} Two classes of points, in a $n$-dimensionnal space, are linearly separable if it exists a hyperplane separating them.
\end{thebibliography}
\end{document}